\documentclass[sigconf]{acmart}
\usepackage{booktabs} % For formal tables
\usepackage{float}
\usepackage{multirow}
\usepackage{graphicx}
\usepackage{array}
\usepackage{xcolor}
\usepackage{soul}
\usepackage{wrapfig}
\usepackage[utf8]{inputenc}
\usepackage{subcaption}
\usepackage{xfrac}

\newcommand{\ignore}[1]{}
\newcolumntype{P}[1]{>{\centering\arraybackslash}p{#1}}
\newcolumntype{M}[1]{>{\centering\arraybackslash}m{#1}}

\AtBeginDocument{%
  }

\copyrightyear{2025}
\acmYear{2025}
\setcopyright{rightsretained}
\acmConference[CHI EA '25]{Extended Abstracts of the CHI Conference on
Human Factors in Computing Systems}{April 26-May 1, 2025}{Yokohama, Japan}
\acmBooktitle{Extended Abstracts of the CHI Conference on Human Factors in
Computing Systems (CHI EA '25), April 26-May 1, 2025, Yokohama,
Japan}\acmDOI{10.1145/3706599.3719991}
\acmISBN{979-8-4007-1395-8/2025/04}

\begin{document}
\setlength{\tabcolsep}{1.5pt}
\title{Making AI-Enhanced Videos: Analyzing Generative AI Use Cases in YouTube Content Creation}


\author{Torin Anderson}
\email{toanderson@clarku.edu}
\orcid{https://orcid.org/0009-0002-4079-4231}
\affiliation{%
  \institution{Clark University}
  \streetaddress{950 Main St.}
  \city{Worcester}
  \state{MA}
  \country{USA}
  \postcode{01610}
}

\author{Shuo Niu}
\email{shniu@clarku.edu}
\orcid{https://orcid.org/0000-0002-8316-4785}
\affiliation{%
  \institution{Clark University}
  \streetaddress{950 Main St.}
  \city{Worcester}
  \state{MA}
  \country{USA}
  \postcode{01610}
}

\renewcommand{\shortauthors}{Shuo Niu et al.}

\begin{abstract}
  Generative AI (GenAI) tools enhance social media video creation by streamlining tasks such as scriptwriting, visual and audio generation, and editing. These tools enable the creation of new content, including text, images, audio, and video, with platforms like ChatGPT and MidJourney becoming increasingly popular among YouTube creators. Despite their growing adoption, knowledge of their specific use cases across the video production process remains limited. This study analyzes 274 YouTube how-to videos to explore GenAI's role in planning, production, editing, and uploading. The findings reveal that YouTubers use GenAI to identify topics, generate scripts, create prompts, and produce visual and audio materials. Additionally, GenAI supports editing tasks like upscaling visuals and reformatting content while also suggesting titles and subtitles. Based on these findings, we discuss future directions for incorporating GenAI to support various video creation tasks.
  
\end{abstract}

\begin{CCSXML}
<ccs2012>
   <concept>
       <concept_id>10003120.10003130.10011762</concept_id>
       <concept_desc>Human-centered computing~Empirical studies in collaborative and social computing</concept_desc>
       <concept_significance>500</concept_significance>
       </concept>
   <concept>
       <concept_id>10003120.10003130.10003131</concept_id>
       <concept_desc>Human-centered computing~Collaborative and social computing theory, concepts and paradigms</concept_desc>
       <concept_significance>500</concept_significance>
       </concept>
 </ccs2012>
\end{CCSXML}

\ccsdesc[500]{Human-centered computing~Empirical studies in collaborative and social computing}
\ccsdesc[500]{Human-centered computing~Collaborative and social computing theory, concepts and paradigms}

\keywords{Generative AI; YouTube; video; content creator}


\maketitle

\section{Introduction}
Generative AI (GenAI) is \textit{``artificial intelligence systems that can create new content, such as text, images, audio, or video, rather than just analyzing or acting on existing data''} \cite{Brizuela2023GenAIReview}. Over the years, the applications of these tools have significantly expanded, including crafting digital content such as creative writing \cite{Lee2024WritingAssist}, short videos \cite{Wang2024ReelFramer}, AI-generated images \cite{Mahdavi2024AIImage}, and music composition \cite{Louie2020Music}. With the emergence of tools like ChatGPT, Sora, MidJourney, and Leonardo.AI, many YouTube content creators have begun incorporating these tools into their video production. On video-sharing platforms \cite{NiuVSPLR}, creators are leveraging GenAI tools to streamline traditionally time-consuming steps, such as writing scripts or creating visual and audio materials \cite{Hoose2024SelfRepresentation, Fancourtbarrierandenablers}. For instance, researchers have observed that YouTubers utilize GenAI tools to create videos on artistic, marketing, knowledge-sharing, and entertainment topics \cite{Lyu2024GenAI}. Despite the rapid adoption of GenAI tools, there remains limited understanding of the specific use cases of these tools in video creation.
\par
Recent research on video creation has explored the challenges inherent in common creative practices and the potential solutions offered by AI-driven approaches \cite{Choi2023Creator, Kim2024ASVG}. During the \textit{Planning Phase}, YouTubers often struggle with finding inspiration and selecting relevant topics. In the \textit{Performance Phase}, creators must dedicate time to recording videos and narrations. Subsequently, the \textit{Editing Phase} requires them to refine their video materials to enhance the overall viewing experience. Finally, during the \textit{Uploading Phase}, creators craft video titles and descriptions aimed at optimizing viewer engagement. These tasks are frequently accompanied by challenges such as generating new ideas, fine-tuning video content, and managing the pressure to produce high-quality, engaging material that attracts audiences and garners popularity \cite{Kim2024ASVG, Choi2023Creator}. While GenAI has demonstrated potential in addressing these challenges \cite{Lyu2024GenAI}, a comprehensive understanding of its use cases within this context remains limited. 
\par
\begin{figure*}[!h]
    \centering
    \includegraphics[width=1\linewidth]{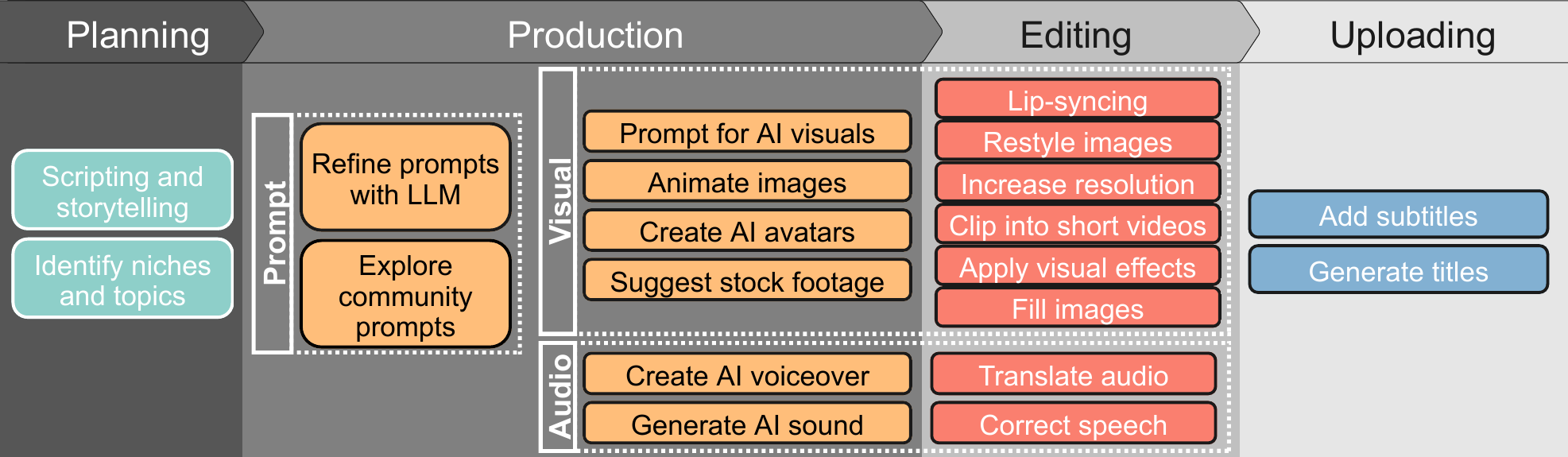}
    \caption{A conceptual framework illustrating GenAI use at different phases of video creation.}
    \label{fig:concept}
\end{figure*}

To bridge this gap, we conduct a preliminary analysis of YouTube how-to videos. How-to videos feature amateur narrators explaining knowledge or teaching skills \cite{Utz2022HowTo}. These videos exemplify YouTube's participatory culture, where peers share skills and enable others to learn, thereby advancing collective knowledge within the community \cite{YouTubeParticipatoryCulture}. Analyzing how-to videos allows us to uncover methods, tips, explanations, descriptions, and conclusions regarding YouTubers' experiences and knowledge of using GenAI tools \cite{Yang2023HowTo}. Our analysis includes the development of a conceptual framework (\autoref{fig:concept}) that illustrates the applications of GenAI in the video creation process. Specifically, we propose a categorization of key GenAI use cases to highlight its roles across various stages of video production, thereby inspiring future designs and the development of GenAI-powered tools to support content creators.
\par
We collected and analyzed 274 videos where creators demonstrated the use cases of GenAI tools in video production. Through thematic analysis, we annotated the use cases of GenAI across the planning, production, editing, and uploading phases. As illustrated in \autoref{fig:concept}, our findings reveal that during the planning phase, YouTubers use GenAI to identify topics and generate video scripts. In the production phase, GenAI is employed to create prompts and produce visual and audio materials. During editing, GenAI tools are used to upscale visuals, resolve video issues, and reformat content. Finally, in the uploading phase, creators rely on GenAI to suggest titles and add subtitles. These applications illustrate how the creator community is leveraging GenAI in practice to streamline video production workflows and reduce the creative labor \cite{Hoose2024SelfRepresentation}.

\section{Method}
The YouTube videos included in our study were retrieved using the YouTube Data API\footnote{\url{https://developers.google.com/youtube/v3}} on March 26, 2024. The search uses the query \textit{``How to (edit OR generate OR create OR make OR use) AI (text OR image OR audio OR video OR animation OR content)''}, which incorporates multiple operations referencing the use of AI and various types of content. This query was set to search videos posted between September 1, 2023, and February 29, 2024, yielding six months of data. The YouTube search process resulted in a total of 3,854 videos. Subsequently, videos were programmatically excluded if they did not have closed captions, were not in English, or had a duration shorter than 470 seconds or longer than 1,372 seconds (considered outliers in duration). After applying these filters, 1,046 videos remained.
\par
To identify GenAI use cases, we randomly selected 200 videos to analyze overall themes. However, we observed that some videos were not how-to videos (e.g., advertisement or news about GenAI), did not involve GenAI, or were unrelated to GenAI in video creation. To ensure relevance, three researchers manually reviewed all videos using three criteria: (1) the video must demonstrate clear steps for using GenAI tools, (2) the tools featured must have generative AI capabilities (e.g., ChatGPT, MidJourney), excluding those focused solely on clipping or editing, and (3) the final output must be a video. A video was kept if at least two of the three annotators agreed on its inclusion. This process resulted in a final dataset of 274 videos for analysis.

\par
To facilitate annotation, we used ChatGPT to segment video transcripts into small clips. The clip is examined individually to mark GenAI use cases. The prompt to segment videos was: \textit{``Split the closed captions into video segments based on coherence between consecutive texts in the closed captions. Each YouTube video segment should include timestamps in the format of `HH:MM:SS <title>'.''} This process resulted in a total of 2,829 video clips for all the videos.

\par
We conducted a thematic analysis \cite{VaismoradiThematicAnalysis} to identify GenAI use cases across the planning, production, editing, and uploading phases of video creation. A total of 150 clips were randomly selected and evenly assigned to three researchers. During the initialization phase, each researcher reviewed the video content and took notes on how GenAI was utilized. To construct use case themes, the researchers compiled all notes and used an affinity diagram to group them around emerging themes. Each identified use case was subsequently categorized into the respective video creation phases. In the rectification phase, two researchers conducted two rounds of annotations for clips of 30 video using the concluded codebook. After each round, the researchers discussed and resolved any disagreements in the annotations.
\par
Finally, two researchers annotated all video clips, selecting all applicable use cases for each clip. The annotations across the four dimensions achieved moderate to substantial agreement (Cohen's kappa can be found in \autoref{tab:cohen}). To consolidate the annotations, a clip was marked as mentioning a use case only if both researchers selected the use case. Subsequently, the annotations of clips belonging to the same videos were merged to determine whether each video contained a specific use case. 

\begin{table}[!h]
    \centering
    \begin{tabular}{|c|c|c|c|c|}
    \hline
        Stages & Planning & Production & Editing & Uploading \\
    \hline
        $\kappa$ & 0.64 & 0.64 & 0.60 & 0.51 \\
    \hline
    \end{tabular}
    \caption{Cohen's cappa of annotation.}
    \label{tab:cohen}
\end{table}

\section{Result}
\begin{figure*}[!h]
    \centering
    \includegraphics[width=1\linewidth]{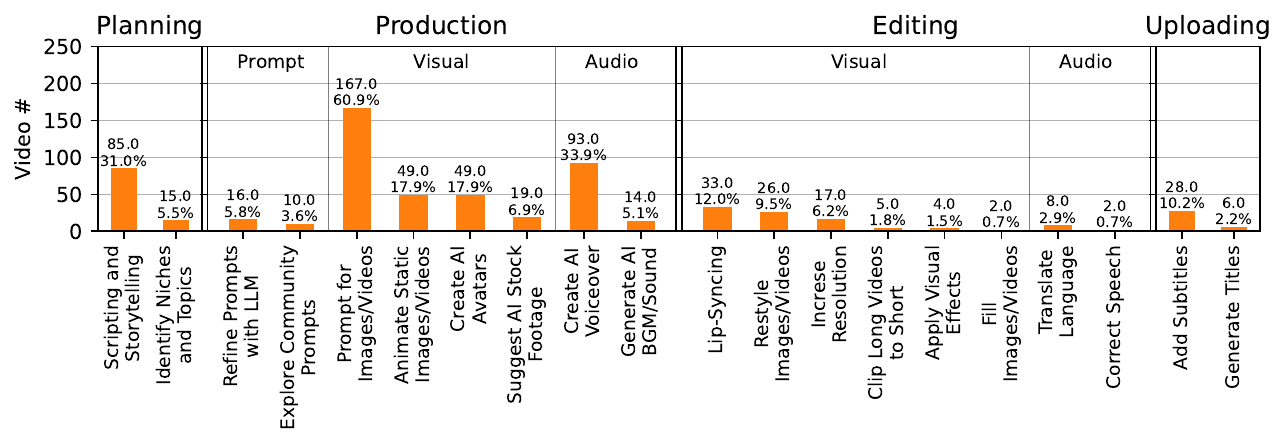}
    \caption{The distribution of videos mentioning each GenAI tool use case across the four stages of video creation.}
    \label{fig:distribution}
\end{figure*}

In this section, we present the use cases along with example videos and the corresponding number of videos introducing each use case. \autoref{fig:distribution} illustrates the distribution of videos featuring GenAI use cases. We categorize the use cases into four phases, primarily based on \cite{Choi2023Creator}. When classifying, we use the following rules. In the \textit{Planning Phase}, content is not yet being created; rather, this phase focuses solely on gathering ideas or topics before taking any further actions. The \textit{Production} and \textit{Editing} phases are distinguished based on whether the use case involves generating entirely new content or modifying existing material. The \textit{Production Phase} encompasses the actualization of video materials, meaning the creation of content for the first time. In contrast, the \textit{Editing Phase} involves utilizing GenAI to modify or update pre-existing content. Finally, the \textit{Uploading Phase} does not involve altering the content itself but instead focuses on generating supplementary information, such as subtitles or titles.

\subsection{Planning Phase}
\begin{figure}[!h]
\centering
    \begin{tabular}{ll}
        \begin{subfigure}[t]{.495\linewidth}
        \includegraphics[width=\linewidth]{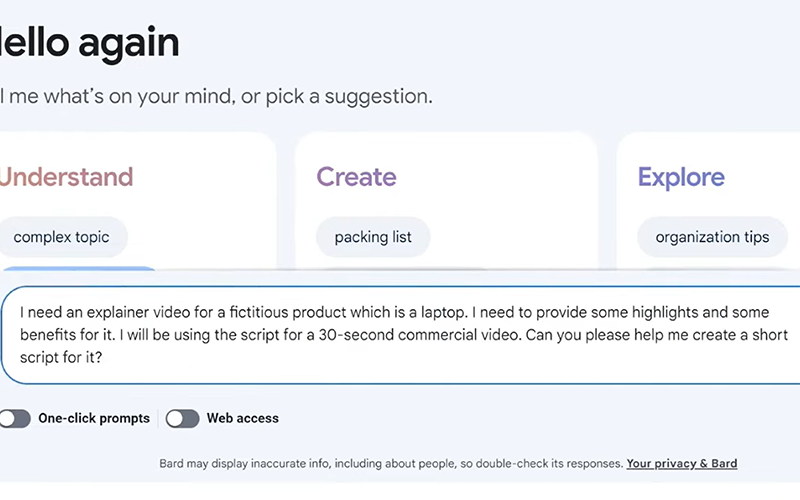} %Ap33gyL90qQ
        \caption{a. Use Google Gemini to create a script for a laptop ads.} 
        \label{fig:scripting}
        \end{subfigure}
        &
        \begin{subfigure}[t]{.495\linewidth}
        \includegraphics[width=\linewidth]{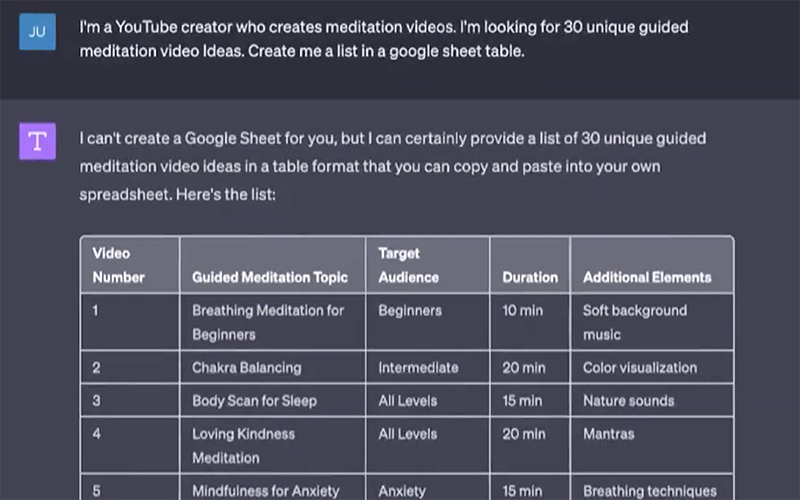} %W8ahuLMQY-I
        \caption{b. Use ChatGPT to explores ideas for meditation videos} 
        \label{fig:niche}
        \end{subfigure}
    \end{tabular}
\caption{Videos with GenAI applications during the Planning phase.}
\label{fig:result1}
\end{figure}

% The \textit{planning} phase involves identifying topics and creating plans for video content. In the how-to videos, YouTubers demonstrate using GenAI tools to discover niche topics and draft video scripts.
\textbf{Scripting and Storytelling.}
A common use case of GenAI in these how-to videos is to create video scripts, as observed in 85 videos (N=85, 31\%). These videos often demonstrate the use of GenAI to generate scripts, video narratives, or storylines. For example, \autoref{fig:scripting} shows a YouTuber using Google Gemini to create a video script for a laptop advertisement. In this case, Gemini generates narration scripts and suggests ideas for the visual design.

\textbf{Identifying Niches and Topics for Videos.}
YouTubers demonstrate the use case of GenAI tools to identify niches or topics for video creation (N=15, 5.5\%). They ask GenAI to suggest a list of topics from which they can choose to create a video. For example, \autoref{fig:niche} illustrates a YouTuber using ChatGPT to generate a table of guided meditation topics, which they later use to create a short-form video.

\subsection{Production Phase}
% The \textit{production} phase is when creators record video materials. In videos created with GenAI, YouTubers leverage these tools to identify effective prompts and generate both visual and audio materials.

\begin{figure}[!h]
\centering
    \begin{tabular}{ll}
        \begin{subfigure}[t]{.495\linewidth}
        \includegraphics[width=\linewidth]{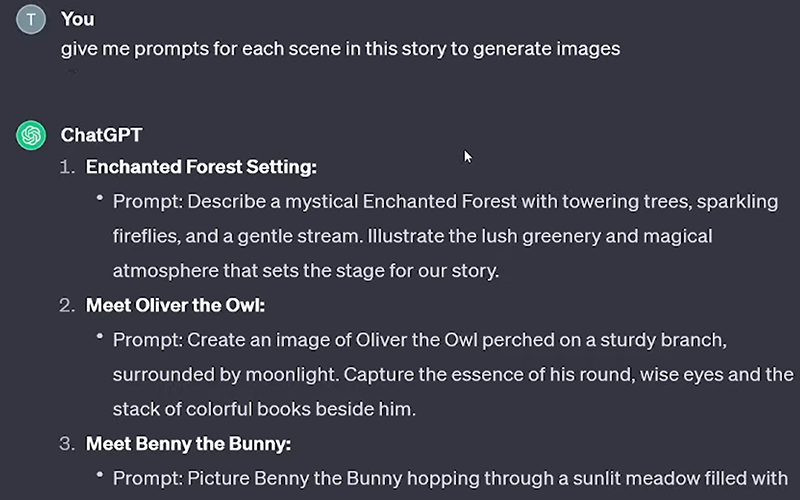} %e2VEkTpOlR8
        \caption{a. Use ChatGPT to create image generation prompts for a fantasy story} 
        \label{fig:prompt}
        \end{subfigure}
        &
        \begin{subfigure}[t]{.495\linewidth}
        \includegraphics[width=\linewidth]{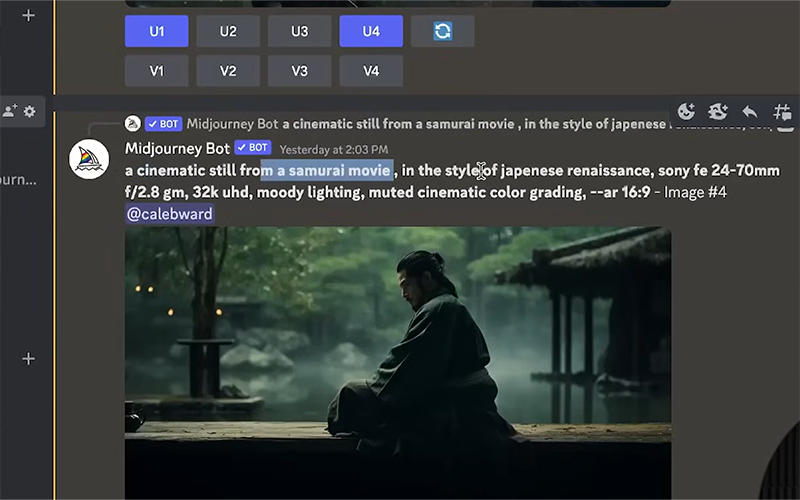} % EgshqVOu1oY
        \caption{b. Cinemtic samurai still generated through Midjourney} 
        \label{fig:community}
        \end{subfigure}
        \\
        \begin{subfigure}[t]{.495\linewidth}
        \includegraphics[width=\linewidth]{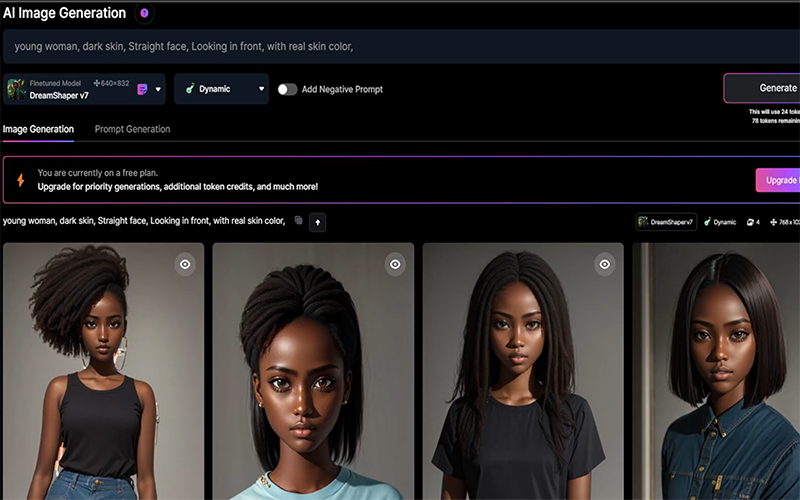} % yKJzEA_uhS4
        \caption{c. Leonardo.AI image generation of young dark skinned woman} 
        \label{fig:image}
        \end{subfigure}
        &
        \begin{subfigure}[t]{.495\linewidth}
        \includegraphics[width=\linewidth]{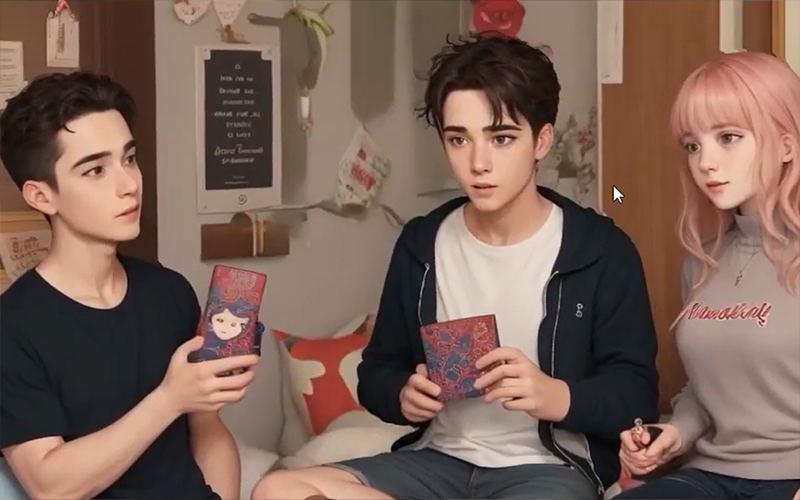} % cslIZ8BUnxk
        \caption{d. Leiapix transforming image of friends to short form video} 
        \label{fig:animate}
        \end{subfigure}
        \\
        \begin{subfigure}[t]{.495\linewidth}
        \includegraphics[width=\linewidth]{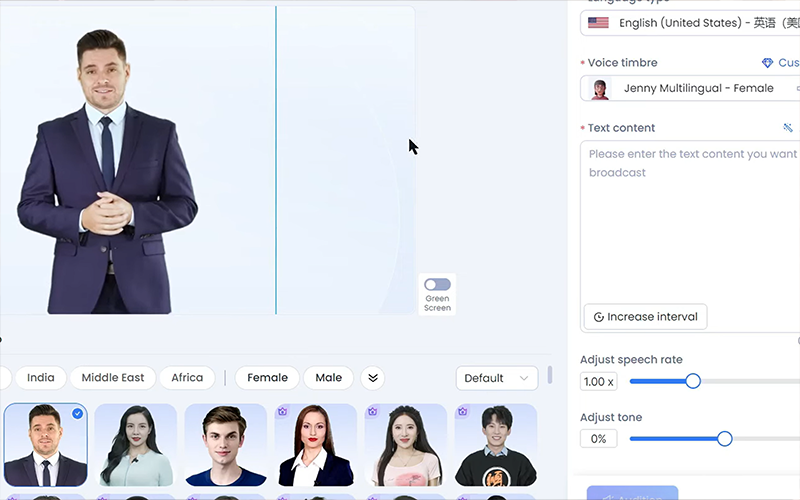} %hl-esiVZSGM
        \caption{e. KreadoAI avatar creation for news presentation style video} 
        \label{fig:avatar}
        \end{subfigure} 
        &
        \begin{subfigure}[t]{.495\linewidth}
        \includegraphics[width=\linewidth]{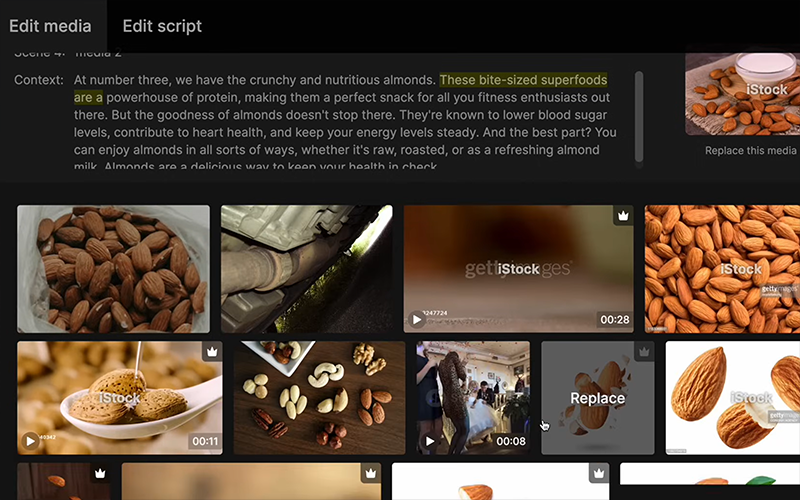} %KwNEwODUAYY
        \caption{f. Invideo AI suggesting generated images of almonds.} 
        \label{fig:stock}
        \end{subfigure}
        \\
        \begin{subfigure}[t]{.495\linewidth}
        \includegraphics[width=\linewidth]{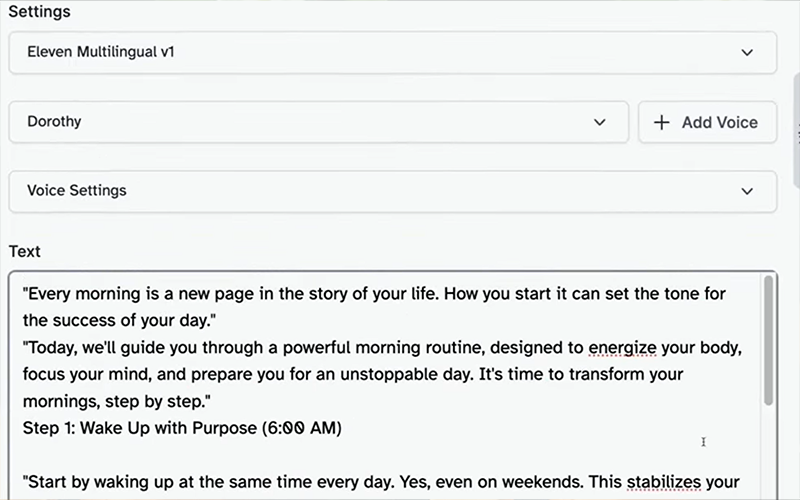} % ZcGl0rb0B7w
        \caption{g. ElevenLabs voiceover generation of female voice for routine video script} 
        \label{fig:voiceover}
        \end{subfigure}
        &
        \begin{subfigure}[t]{.495\linewidth}
        \includegraphics[width=\linewidth]{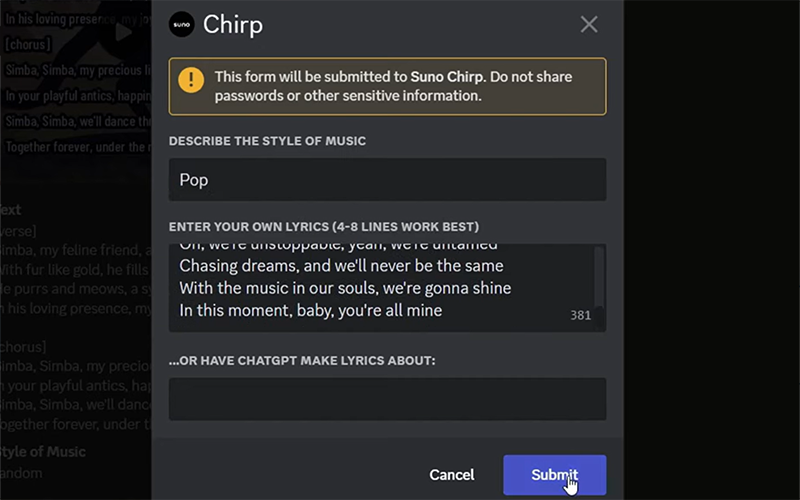} %cvRJ_izhs28
        \caption{h. Suno AI music generation to create a pop song} 
        \label{fig:music}
        \end{subfigure}
        
    \end{tabular}
\caption{Videos with GenAI applications during the Production phase.}
\label{fig:result2}
\end{figure}

\textbf{Refining Prompts with LLM.}
A method for obtaining GenAI prompts is by directly asking GenAI tools to refine the prompt (N=16, 5.8\%). To improve prompts and generate high-quality AI materials, YouTubers recommend using LLM tools, such as ChatGPT, to create prompts tailored to specific video content, such as image generation or voiceover production. For instance, \autoref{fig:prompt} illustrates a YouTuber using ChatGPT to craft detailed prompts for an AI image-generation tool. These prompts include specific scenes and detailed descriptions designed to produce high-quality images suitable for a children's YouTube video.

\textbf{Exploring Community-Created Prompts.}
In how-to videos, YouTubers recommend use and customize the community-created AI prompts (N=10, 3.6\%). These videos suggest reviewing successful prompts for AI image or video generation tools to learn how to craft them. For example, \autoref{fig:community} showcases a YouTuber analyzing an image generated by MidJourney along with its corresponding prompt. The YouTuber highlights the keywords used in the prompt to create ideal AI images.

\textbf{Prompting for Images/Videos.}
In our videos, YouTubers commonly showcase how to use GenAI tools to generate AI images and videos (N=167, 60.9\%). Creators often use tools such as Leonardo.AI, Midjourney, and Canva to produce images or short video clips, which are video materials to be integrated into their videos. For instance, as shown in \autoref{fig:image}, a YouTuber utilizes Leonardo.AI to create a portrait image of a young, dark-skinned woman. In this case, the YouTuber uses the generated image as an avatar for their video.

\textbf{Animating Static Images to Videos.}
To obtain video clips, YouTubers demonstrate the use case of GenAI tools to animate static images into videos (N=49, 17.9\%). These videos leverage GenAI tools to transform an image into short-form videos that can used as standalone video content or be combined into a longer video. For instance, \autoref{fig:animate} highlights a YouTuber using Leiapix to create a short video from a cartoon-style image of a group of friends. This short video is then incorporated into a longer video featuring several such short AI video clips.

\textbf{Creating AI Avatars.}
To create videos featuring talking heads, YouTubers introduce various tools to generate human-like avatars that can read a video script (N=49, 17.9\%). For example, \autoref{fig:avatar} shows a YouTuber using KreadoAI to select an avatar from a list of options. The YouTuber chooses a professionally dressed avatar for use in a news video.

\textbf{Suggesting AI Stock Footage.}
The GenAI community use video AI tools to look for and select stock footage based on the texts in their video scripts (N=19, 6.9\%). These videos utilize visual editing tools, such as Invideo AI, to replace sections of the video with AI-selected stock images and videos. For instance, \autoref{fig:stock} features a YouTuber using Invideo AI to identify various stock footages of almonds to be presented in a video.

\textbf{Creating AI Voiceover.}
To create soundtracks, YouTubers recommend using AI voiceovers for video narration (N=93, 33.9\%). These videos use text-to-speech tools to generate audio that narrates the given text script. For example, \autoref{fig:voiceover} illustrates a YouTuber using ElevenLabs to produce a female voiceover for a script outlining healthy morning routines. 

\textbf{Generating AI BGM/Sound Effects.}
GenAI tools are also used to create music/sound effects (N=14, 5.1\%). For example, in \autoref{fig:music}, a YouTuber demonstrates using Suno AI to create a pop song based on the YouTuber-provided lyrics.

\subsection{Editing Phase.}
% The \textit{editing} phase is when creators further improve the video materials to improve the quality of the video.

\begin{figure}[!h]
\centering
    \begin{tabular}{ll}
        \begin{subfigure}[t]{.495\linewidth}
        \includegraphics[width=\linewidth]{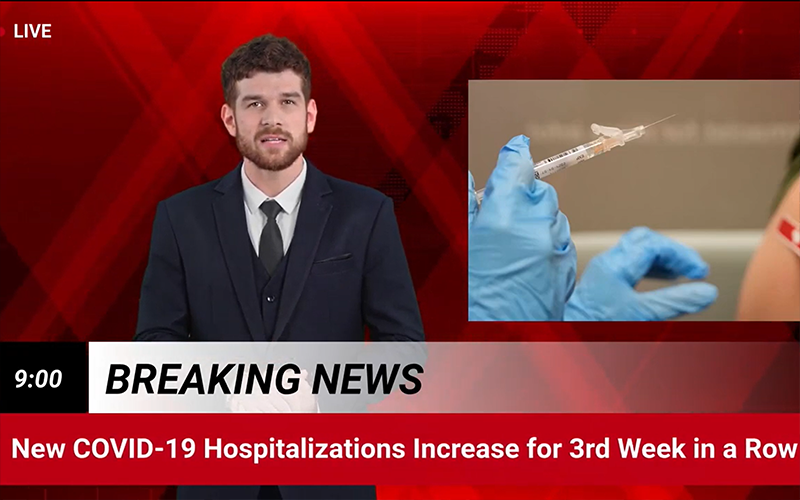} %Q4qJduJyp64
        \caption{a. Vidnoz AI lip syncing news broadcaster avatar lip's to audio recording} 
        \label{fig:lip-sync}
        \end{subfigure}
        &        
        \begin{subfigure}[t]{.495\linewidth}
        \includegraphics[width=\linewidth]{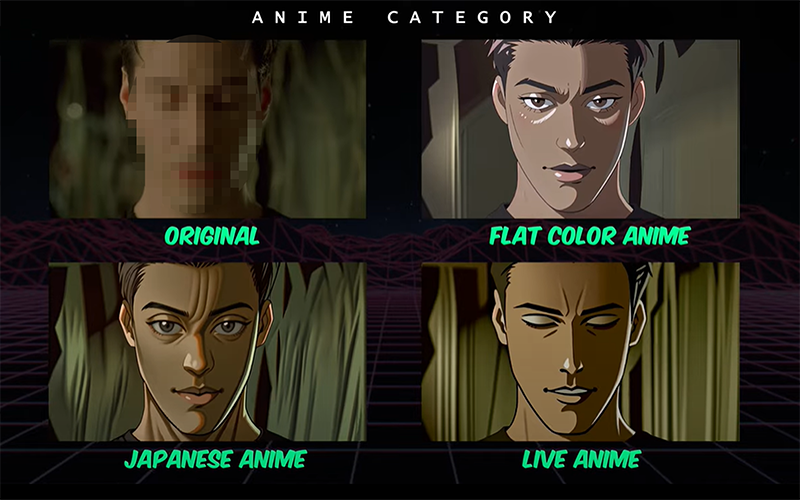} % FyOmfk3qZUA
        \caption{b. DomoAI restyling a video into multiple different styles} 
        \label{fig:restyle}
        \end{subfigure}
        \\
        \begin{subfigure}[t]{.495\linewidth}
        \includegraphics[width=\linewidth]{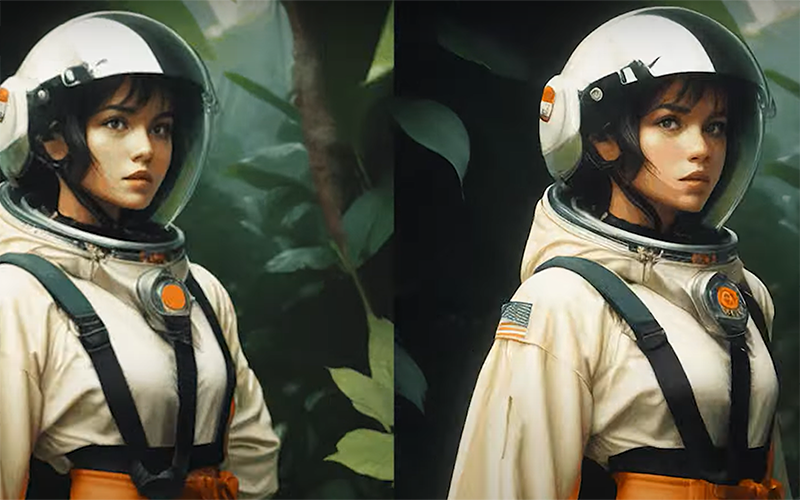} %kmZ5S2X55fU
        \caption{c. ComfyUI upscaling an animated portrait video of a woman} 
        \label{fig:resolution}
        \end{subfigure}
        &
        \begin{subfigure}[t]{.495\linewidth}
        \includegraphics[width=\linewidth]{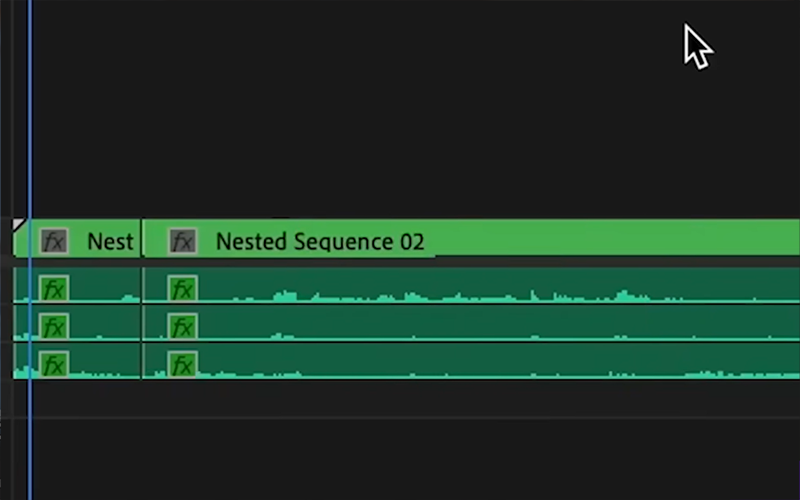} %KlnxT2EhkHo
        \caption{d. Autopod breaking a podcast into shorter clips} 
        \label{fig:clipping}
        \end{subfigure}
        \\
        \begin{subfigure}[t]{.495\linewidth}
        \includegraphics[width=\linewidth]{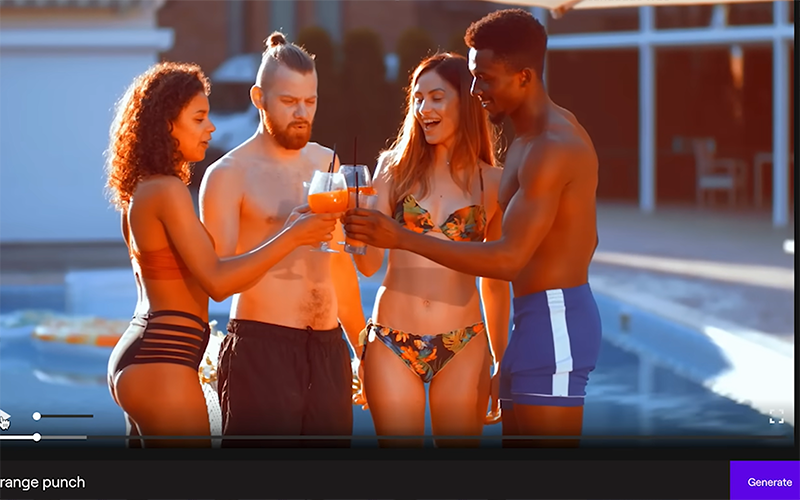} %KlnxT2EhkHo
        \caption{e. Lut adding a filter over an image taken by a pool} 
        \label{fig:background}
        \end{subfigure}
        &
        \begin{subfigure}[t]{.495\linewidth}
        \includegraphics[width=\linewidth]{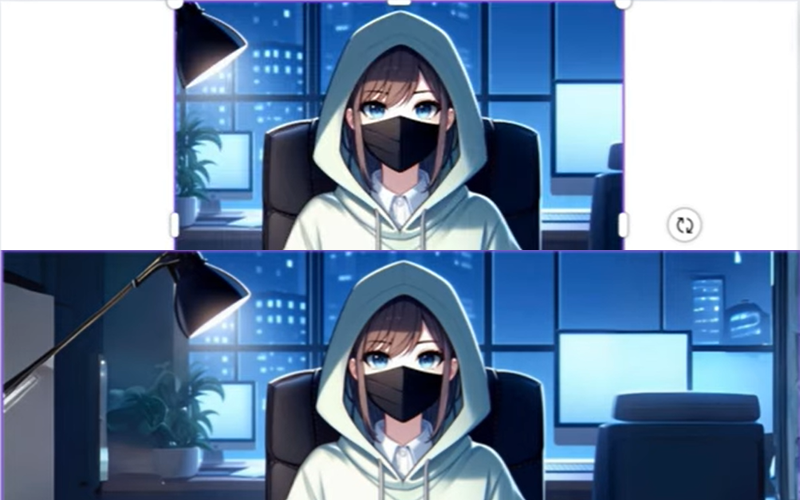} %eK2ehGSmPx4
        \caption{f. Canva AI filling an image on the sides}
        \label{fig:fill}
        \end{subfigure}
        \\
        \begin{subfigure}[t]{.495\linewidth}
        \includegraphics[width=\linewidth]{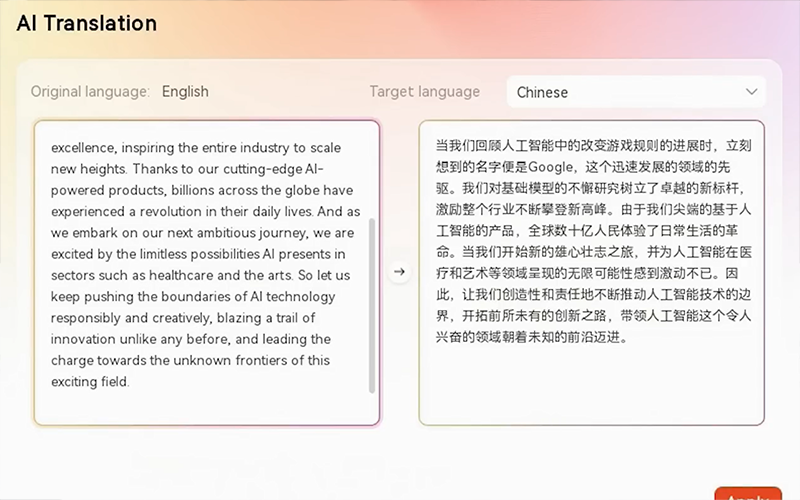}
        \caption{g. Wondershare Virbo translating a script into Chinese} %U86cC0s7zj8
        \label{fig:translate}
        \end{subfigure}
        &
        \begin{subfigure}[t]{.495\linewidth}
        \includegraphics[width=\linewidth]{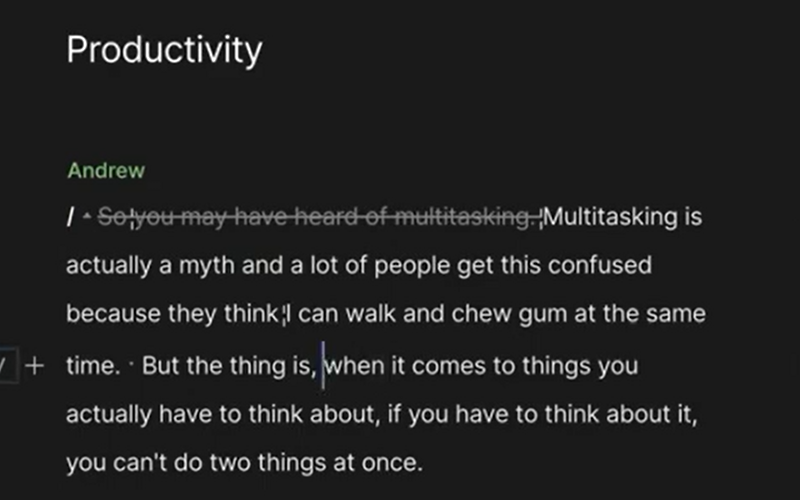} %r1ZJ0qFQkxs
        \caption{f. Descript cutting unnessary script from a demo style video} 
        \label{fig:enhance}
        \end{subfigure}
    \end{tabular}
\caption{Videos with GenAI applications during the Editing phase.}
\label{fig:result3}
\end{figure}

\textbf{Lip-Syncing.}
YouTubers also demonstrate how to use GenAI for lip-syncing AI avatars or photos of real persons (N=33.0, 12.0\%). These videos show using GenAI tools to sync the avatar's lips with the speech from an uploaded script. For example, \autoref{fig:lip-sync} shows a YouTuber utilizing Vidnoz AI to create a news-style video with an AI news presenter. Vidnoz automatically lip-syncs the avatar with the uploaded script.

\textbf{Restyling Images/Videos.}
GenAI tools are used to restyle images and videos (N=26, 9.5\%). The how-to Youtubers utilize image GenAI tools to transform image styles or alter the atmosphere of visual content. For example, \autoref{fig:restyle} features a YouTuber using DomoAI to restyle a video, showcasing the conversion of a movie clip into flat color, Japanese, and live anime styles.

\textbf{Increasing Resolution.}
The YouTubers that specialize in GenAI how-tos also use GenAI tools to enhance the resolution of videos and images (N=17, 9.5\%). For instance, \autoref{fig:resolution} shows a YouTuber using ComfyUI to upscale an animated video by improving its resolution.

\textbf{Clipping a Long Video to Short Videos.}
GenAI tools are also used to automatically clip long videos and create short-form videos (N=5, 1.8\%). These short-form clips can then be uploaded to platforms like TikTok. For example, in \autoref{fig:clipping}, a YouTuber using Autopod to automatically break up a podcast video. The YouTuber break down the podcast into parts of similar length that can be posted separately. The tool automatically identifies important sections based on a ``virality score.''

\textbf{Applying Visual Effects.}
Some how-to videos use GenAI tools to add visual effects or modify image backgrounds to better align with the video theme (N=4, 1.5\%). These creators upload figures and instruct GenAI tools to alter the backgrounds or apply video filter effects. For instance, \autoref{fig:background} features a YouTuber using Lut to add an orange tint to a photo of friends.

\textbf{Filling Images/Videos.}
YouTubers recommend using GenAI to fill areas of images (N=2, 0.7\%). These videos utilize AI visual editing tools to expand or fill empty spaces within an image. For instance, \autoref{fig:fill} shows a YouTuber using Canva to generate an animation-style image and then expand and fill the sides to create a wider picture to fit the video's aspect ratio.

\textbf{Translating Language.}
YouTubers in their how-to videos leverage GenAI tools to translate the languages of their videos (N=8, 2.9\%). These videos often feature avatars, and YouTubers aim to convert speech from one language to another. For instance, in \autoref{fig:translate}, a YouTuber demonstrates Wondershare Virbo, showing how a translation from English to Chinese works by using a desired script. This newly translated script is automatically lip-synced afterwards.

\textbf{Correcting Speech.}
In the how to videos, YouTubers demonstrate the use case of GenAI to correct speech errors in videos (N=2, 0.7\%). They upload an audio or video recording and utilize GenAI tools to eliminate pauses in their speech. As shown in \autoref{fig:enhance}, a YouTuber demonstrates how Descript removes unnecessary script sections. These sections are also seamlessly edited out from the video.

\subsection{Uploading Phase}

\begin{figure}[!h]
\centering
    \begin{tabular}{ll}
        \begin{subfigure}[t]{.495\linewidth}
        \includegraphics[width=\linewidth]{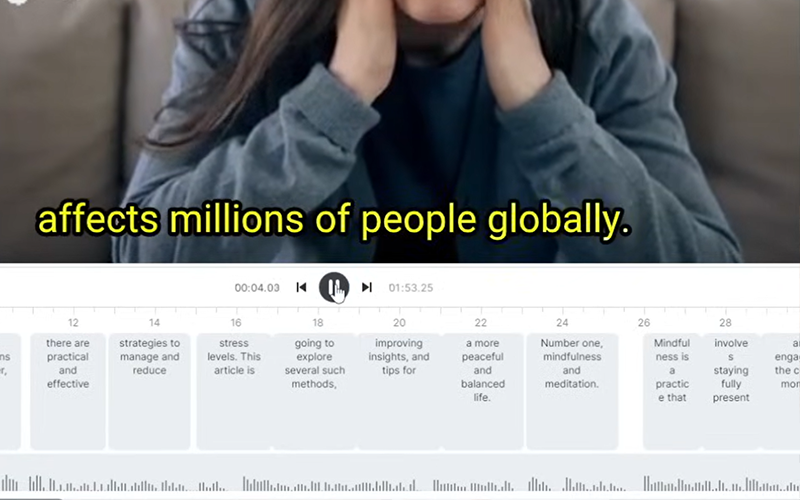} %J776-wcVY4c
        \caption{a. Using wave.video to transcribe and create subtitles for mindfullness video}
        \label{fig:subtitle}
        \end{subfigure}
        &
        \begin{subfigure}[t]{.495\linewidth}
        \includegraphics[width=\linewidth]{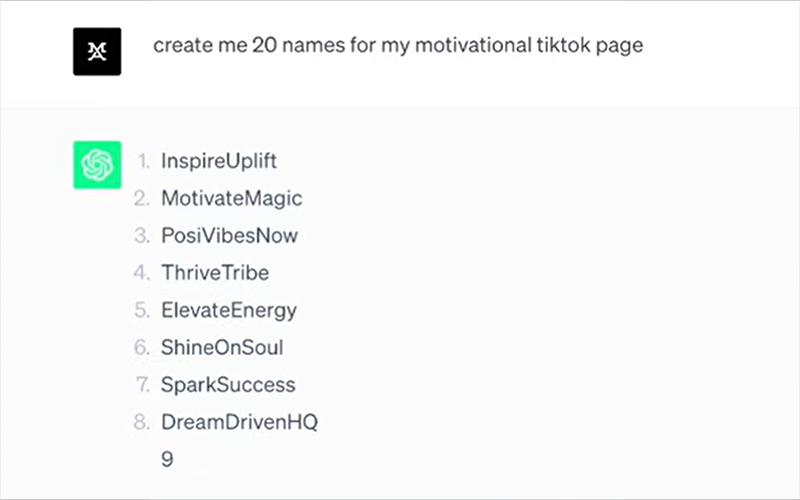} %Usg3PIiVKfQ
        \caption{b. Use ChatGPT to generate different names for a motivation Tiktok page} 
        \label{fig:name}
        \end{subfigure}
    \end{tabular}
\caption{Videos with GenAI applications during the Uploading phase.}
\label{fig:result4}
\end{figure}

\textbf{Adding Subtitles.}
In the final phase, how-to videos use GenAI to help with video subtitles (N=28, 10.2\%). These videos often utilize GenAI to process video content and create captions. For example, \autoref{fig:subtitle} shows a YouTuber using wave.video to transcribe and automatically add subtitles to a video.
        
\textbf{Generating Titles.}
Some how-to videos also use GenAI tools to suggest video titles (N=6, 2.2\%). They use ChatGPT to create video titles or channel names before publishing the videos. For example, \autoref{fig:name} shows a YouTuber using ChatGPT to generate names for a TikTok account focused on motivational content.

\section{Discussion}
Based on our results, the use cases can be summarized within a conceptual framework, as presented in \autoref{fig:concept}. While these use cases were derived from an analysis of YouTube how-to videos on generative AI, the identified stages can be broadly applied to creative processes for video creation on other platforms, such as TikTok and Instagram. In the following sections, we discuss future directions for enhancing and supporting the use of GenAI by video content creators. Based on these GenAI usage patterns, we outline key questions to be considered and examined in HCI research.

\subsection{A Support or Hindrance to Ideation and Creativity}
When planning video content, YouTubers leverage GenAI to identify topics and generate scripts and storylines. The use of GenAI introduces key considerations regarding its impact on creative labor. Prior research has highlighted that sustaining creative labor is a significant challenge for YouTube content creators \cite{Hoose2024SelfRepresentation}. In the content creation industry, demonstrating creativity and active participation has traditionally been essential for user engagement \cite{Burgess2006}. Our findings suggest that GenAI can play a role in this process, potentially blurring the boundary between human and AI-generated creativity.
\par
The HCI community must critically examine the role of generative AI in ideation and creative tasks, focusing on its support mechanisms, challenges, and strategies for mitigating its limitations. Future research should explore legitimate and organic approaches to integrating GenAI tools in ways that inspire creators while preventing over-reliance on AI, as excessive AI-generated content may undermine originality and limit creativity. Assessing users' perceptions of GenAI-identified video topics and scripts is essential for developing practical guidelines on effectively incorporating GenAI into scriptwriting and narration, thereby enhancing its role in video creativity. Given GenAI's tendency to produce repetitive or similar content \cite{hwang2023aligninglanguagemodelsuser}, it is crucial to investigate strategies that enable YouTubers to leverage GenAI effectively while maintaining distinctiveness and engagement.

\subsection{Prompt Creation as a Critical Skill}
Our analysis of GenAI in the production and editing phases highlights its diverse utility across various types of AI-generated content. YouTubers refine community-shared GenAI prompts and use LLMs to generate high-quality visual and audio materials. These use cases suggest that the design of GenAI tools should consider the need for prompt support and the evolving role of prompts in future user-generated videos.

\par
With the growing use of GenAI, prompting skills may become a critical component of content creation. Similar to other skills in video production \cite{YouTubeParticipatoryCulture}, tips and knowledge on effective prompting may emerge as a new form of expertise exchanged within the content creator community. Reflecting on this trend, there is an opportunity to design scaffolding mechanisms for content creation using GenAI tools. To support video production, designers could integrate GenAI functions into video editing tools, facilitating access to community-shared prompts and leveraging LLM tools to enhance prompt effectiveness. Such designs should ensure that the generated content aligns with the creator's video concept. Another design opportunity is to support creators with disabilities \cite{BorgosYouTubeDisability, Niu2024Disability}. Future research should explore effective GenAI-based support mechanisms that enable creators with diverse disabilities to overcome challenges in producing digital content. For instance, creators experiencing conditions such as anxiety or depression often face barriers to video production \cite{Fancourtbarrierandenablers}. GenAI applications can be designed to assist these creators by streamlining video editing processes or correcting speech glitches \cite{Arriagadayouneedatleast, Yubarrierstoindustry}.

\subsection{Originality and Authenticity of Multimodal AI Content}

In our data, we noticed the multimodality use of GenAI materials being integrated together. For visuals, GenAI creates AI images, videos, animations, talking avatars, and suggests stock footage. For audio, it generates voiceovers, music, and sound effects. In editing, GenAI upscales content, enhances resolution, applies effects, fills in details, enables lip-syncing, and converts long-form videos into short clips. It also improves audio by correcting speech and translating languages. These use cases suggest that the design of GenAI tools should account for the diversity of styles and the impact on the authenticity of the content.

\par

The multimodal content produced by GenAI, encompassing visual and audio elements, raises new research questions regarding the effective use and integration of content generated by different GenAI tools. There is still a lack of practical guidelines on how different AI-generated content affects video outcomes. For example, should creators use an AI avatar or their own face? Should they rely on generative AI to fill or fix images, or should they replace them with AI-generated stock footage? Does using AI voiceovers impact audience engagement? Therefore, it is essential to explore the factors influencing creators' adoption of various GenAI tools, the specific tasks they use them for, and the perceived value of GenAI-generated content. Additionally, it is critical to examine users' acceptance and rejection of different AI applications.

\par

In the content creation industry, certain tasks have traditionally been time-consuming, such as producing high-quality video visuals, recording talk-to-camera footage, or creating engaging background music. However, it remains unclear whether GenAI-generated content, when prompted or edited by creators, is still considered original. Prior studies indicate that human-created art is often preferred over AI-generated art \cite{Bellaichehumansvsai}. A central question in this context is the originality and authenticity of AI-generated content in representing the creator. While GenAI-generated video and audio content may convey a sense of being ``made by AI,'' it is essential to assess whether such content leads to a loss of connection with the audience and diminishes organic social interaction within the community.

\subsection{Visibility and Transparency of GenAI-Generated Content}
Before uploading, YouTubers use GenAI to automatically generate subtitles and create video titles. Adding subtitles is crucial for enhancing the viewer experience \cite{Gemsbacher2015Caption}, while effective video titles can increase a video's visibility on social media platforms \cite{Jiang2014ViralVideo}. The use of GenAI for these video completion tasks suggests that future designs could integrate GenAI tools to streamline these processes for creators. 

\par
On video-sharing platforms, recommendation algorithms play a critical role in determining creators' visibility \cite{NiuVSPLR}. Creators are also deeply concerned about their algorithmic visibility on these platforms \cite{Bishop2019Visibility}. When YouTubers adopt AI-generated content to ``improve'' aspects such as presentation and meta-information (e.g., titles and descriptions), they must consider whether such content may be shadow-banned or demonetized by the platform in accordance with community guidelines regarding AI content\footnote{\url{https://blog.youtube/inside-youtube/our-approach-to-responsible-ai-innovation/}}. Researchers should investigate how AI-generated or AI-modified content must be labeled and disclosed, as well as examine the impact of these factors on the algorithm’s ability to promote or demote videos.

\section{Future Work}
The follow-up to this work will focus on two key directions. First, as highlighted by the significant use cases identified in our data, it is essential to examine and evaluate the design of new GenAI tools to support these video creation activities, including scriptwriting, creating voiceovers, and facilitating prompt formation. Additionally, HCI research should explore the practices and acceptance of AI-enhanced videos among creators and viewers to gain a deeper understanding of potential ethical concerns and their impact on the quality of YouTube videos.

\bibliographystyle{ACM-Reference-Format}
\bibliography{references, references_ta}

\end{document}